\documentclass[3p]{elsarticle}

\usepackage{hyperref}
\usepackage{xspace}
\usepackage{hyphenat}
\usepackage{amsfonts}
\usepackage{ragged2e}
\usepackage{url}
\usepackage[table,xcdraw]{xcolor}
\usepackage{color}

\def \dataset{{\footnotesize{\texttt{ULCER\_SET}}}\xspace}
\def\system {\textit{QTDU}\xspace}
\def\S {\mathcal{S}\xspace}
\def\I {\mathcal{I}\xspace}

\journal{Computer Methods and Programs in Biomedicine}

\begin{document}

\begin{frontmatter}

\title{A superpixel-driven deep learning approach for the analysis of dermatological wounds\tnoteref{mytitlenote}}
\tnotetext[mytitlenote]{The authors thank FAPESP,  FAPERJ, and CNPq for their financial support. Formal publication DOI: https://doi.org/10.1016/j.cmpb.2019.105079. $\textsuperscript{\textcopyright}$ 2019. This manuscript version is made available under the CC-BY-NC-ND 4.0 license http://creativecommons.org/licenses/by-nc-nd/4.0/}

%% Group authors per affiliation:
\author{Gustavo Blanco$^a$, Agma J. M. Traina$^{a,*}$, Caetano Traina Jr.$^a$, and Paulo M. Azevedo-Marques$^b$}
\address{Institute of Mathematics and Computer Sciences -- ICMC/USP -- Brazil$^a$ \\ Ribeir\~{a}o Preto Medical School - HCFMRP/USP -- Brazil$^b$}
\cortext[mycorrespondingauthor]{Corresponding author. Email: \{agma, bedo\}@icmc.usp.br.}

\author{Ana E. S. Jorge$^c$}
\address{Department of Physical Therapy -- DFisio/UFSCar -- Brazil$^c$}

\author{Daniel de Oliveira$^d$, and Marcos V. N. Bedo$^e$}
\address{Institute of Computing -- IC/UFF -- Brazil$^d$ \\ Fluminense Northwest Institute -- INFES/UFF -- Brazil$^e$}

\begin{abstract}
\textit{Background.} The image-based identification of distinct tissues within dermatological wounds enhances patients' care since it requires no intrusive evaluations.
This manuscript presents an approach, we named \system, that combines deep learning models with su\-per\-pixel-driv\-en segmentation methods for assessing the \underline{q}uality of \underline{t}issues from \underline{d}ermatological \underline{u}lcers.\\
\textit{Method.} \system consists of a three-stage pipeline for the obtaining of ulcer segmentation, tissues' labeling, and wounded area quantification.
We set up our approach by using a real and annotated set of dermatological ulcers for training several deep learning models to the identification of ulcered superpixels.\\
\textit{Results.} Empirical evaluations on $179{,}572$ superpixels divided into four classes showed \system accurately spot wounded tissues (AUC = $0.986$, sensitivity = $0.97$, and specificity = $0.974$) and outperformed machine-learning approaches in up to $8.2\%$ regarding F1-Score through fine-tuning of a ResNet-based model.
Last, but not least, experimental evaluations also showed \system correctly quantified wounded tissue areas within a $0.089$ Mean Absolute Error ratio.\\
\textit{Conclusions.}
Results indicate \system effectiveness for both tissue segmentation and wounded area quantification tasks.
When compared to existing machine-learning approaches, the combination of superpixels and deep learning models outperformed the competitors within strong significant levels.

\end{abstract}


\end{frontmatter}

%\linenumbers

\section{Introduction}\label{sec:introduction}

The growing number of different devices that support medical image acquisition and the ever-decreasing costs for storing such images have given rise to new and larger image databases in several medical centers way beyond the radiology room~\cite{Gillies2015}.
For example, protocols for collecting images of lower limb ulcers by using low-cost smartphones in controlled environments have shown great potential to be included in the clinical workflow as additional information that supports physicians' analyses~\cite{Yu2017,Deserno2018}.
Moreover, such images can be automatically evaluated by Computer-Aided Diagnosis (CAD) tools, or even used for the searching of massive databases through content-only queries, as in Content-Based Image Retrieval (CBIR) applications.
In both CAD and CBIR cases, the detection of abnormalities requires the extraction of patterns from images, while a decision-making strategy is necessary for juxtaposing new images to those in the database~\cite{Blanco2016,Zahia2018}.  

Since dermatological lesions are routinely diagnosed by biopsies and surrounding skin aspects, ulcers can be computationally characterized by particular types of tissues (and their areas) within the wounded region~\cite{Litjens2017,Seixas2015}.
For instance, Mukherjee \textit{et al.}~\cite{Mukherjee2014} proposed a five-color classification model and applied a color-based low-level extractor further labeled by a Support-Vector Machine (SVM) strategy at an $87.61\%$ hit ratio.
Such idea of concatenating feature extraction and classification is found at the core of most wound segmentation strategies, as in the study of Kavitha \textit{et al.}~\cite{Kavitha2017} that evaluated leg ulcerations by extracting patterns based on local spectral histograms to be labeled by a Multi-Layer Perceptron (MLP) classifier with $87.05\%$ accuracy.
Analogously, Pereyra \textit{et al.}~\cite{Pereyra2014} discussed the use of color descriptors and an Instance-based Learning (IbL) classifier with a 61.7\% hit ratio, whereas Veredas \textit{et al.}~\cite{Veredas2010} suggested the use of texture descriptors and an MLP classifier with 84.84\% accuracy.

Blanco \textit{et al.}~\cite{Blanco2016} and Chino \textit{et al.}~\cite{Chino2018} followed a slightly different premise for finding proper similarity measures and comparison criteria for dermatological wounds.
Their approaches are based on a divide-and-conquer strategy, in which an ulcer is segmented with the support of \textit{superpixel} construction methods~\cite{Achanta2012}.
Such methods are employed for splitting an image into several pieces of ulcered tissues with well-defined borders to be described by feature engineering methods, as MPEG-7 descriptors~\cite{Blanco2016} and Bag-of-Signatures~\cite{Chino2018}. 
The features are labeled by the RandomForest classifier, which segments dermatological wounds with $89.87\%$ accuracy~\cite{Chino2018}.
Those classifier-driven segmentation approaches also provide the basis for measurements of the size and stage of the wound~\cite{Blanco2016,Mukherjee2014,Kavitha2017,Pereyra2014,Veredas2010,Dorileo2010,Khalil2019}.

Recently, deep-learning (DL) models have been successfully applied to specific tissue segmentation problems, such as skin cancer and melanoma characterization~\cite{Esteva2017,Yuexiang2018,Wang2015}.
They usually rely on convolutional neural networks (CNNs) for combining both feature engineering and data classification into a single package, which eliminates the need for data extraction.
DL models distinguish themselves regarding the topology of the underlying CNN, \textit{e.g.}, VGG~\cite{Simonyan2014}, AlexNet~\cite{Russakovsky2015}, Resnet~\cite{He2016} or InceptionV3~\cite{Szegedy2016}, and the learning algorithm, which can be either end-to-end training or transfer-learning.
For instance, Goyal \textit{et al.}~\cite{Goyal2018} proposed a transfer-learning DL model for diabetic foot segmentation with $92.5\%$ accuracy, while Nejati \textit{et al.}~\cite{Nejati2018} combined AlexNet and SVM for wound classification at an 86.40\% hit ratio.
Analogously, Zahia \textit{et al.}~\cite{Zahia2018} implemented a divide-and-conquer method that split pressure ulcers into fixed size regions to be labeled by a CNN at a $91.3\%$ Dice Coefficient ratio.
The reviewed approaches point out that the challenges of applying DL models for the detection of wounded tissues are related to 
\textit{(i)}~finding the most suitable CNN architecture, 
\textit{(ii)}~determining the applicability of transfer-learning methods from trained CNNs, and
\textit{(iii)}~expert-burden or biopsy-based labeling of a massive amount of images~\cite{Deserno2018,Wang2015,Shin2016}.

This manuscript presents the \system approach for the assessment of the \underline{q}uality of \underline{t}issues from \underline{d}er\-ma\-tol\-o\-gical \underline{u}lcers.
\system integrates previous literature efforts into DL models towards an accurate segmentation of ulcer tissues in lower limbs by adopting a \textit{divide-and-conquer} strategy through superpixels that define the wounded tissue candidates to be learned by DL models.
The idea is only a few images must be labeled by experts, from which hundreds of thousands of superpixels can be automatically extracted and used for the training of CNNs.
A large set of experiments was designed for finding both the most suitable CNN architecture and the transfer-learning method that fit the purpose of segmenting dermatological ulcers, and \system was compared to previous segmentation efforts in a testbed based on superpixels labeled by human experts.
The results showed our approach outperformed the competitors by significant margins and correctly quantified the pixelwise wounded area.
Accordingly, the contributions of the manuscript are as follows:

\begin{enumerate}

\item The design of an approach that integrates superpixel methods into DL models for the identification of distinct tissues within ulcered areas. 
The strategy, coined \system, enables both dermatological ulcer segmentation and pixelwise area quantification, and

\item Experimental indications of \system most suitable parameters: \textit{(i)}~raw SLIC superpixels, and \textit{(ii)}~ResNet DL model with the addition of six new layers as part of transfer-learning.\\

\end{enumerate}

The remainder of the manuscript is organized as follows:
Section~\ref{sec:background} discusses preliminaries concepts on wound segmentation;
Section~\ref{sec:method} addresses materials and methods, and a \system description;
Section~\ref{sec:experiments} focuses on experimental evaluations.
Finally, Section~\ref{sec:conclusion} provides the conclusions.

\section{Preliminaries}\label{sec:background}

While the segmentation of wounded tissues from photographic images of dermatological ulcers has been discussed from distinct perspectives~\cite{Deserno2018,Mukherjee2014,Nejati2018}, most of the current approaches perform a three-stage pipeline for the labeling of wounds towards specific beacons and markers~\cite{Pereyra2014,Veredas2010}.
Such a pipeline is composed of 
\textit{(i)}~\underline{region segmentation}, which aims to remove image noise and delimit region boundaries, 
\textit{(ii)}~\underline{feature extraction}, which represents (parts of) an image in a multidimensional space, and 
\textit{(iii)}~\underline{data classi-} \underline{fication}, which assigns a label to each image representation.

Unlike existing approaches, our premise is raw image segmentation by superpixels can be combined with DL models to improve the current wound segmentation pipeline and enhance tissue labeling since that combination removes the need for low-level feature engineering.
Moreover, advances in distinct yet related areas, such as melanoma detection~\cite{Litjens2017,Han2018,Youssef2018}, can be incorporated into the models through transfer-learning methods.
The following paragraphs describe the concepts required by our method.\\

\noindent
\textbf{Notation.} 
An image $I$ is a structured set of pixels $P_i$, such that $I = \{P_i~|~0 \leq i \leq n\}$, where $n$ is the total number of pixels of $I$. 
A pixel $P_i$ is a triple $P_i= (r_i, g_i, b_i)$, where $r_i$, $g_i$, and $b_i$ represent the pixel intensity in the RGB color space~\cite{Sonka2014}. 
Analogously, a superpixel $S$ corresponds to a structured subset $S \subseteq I,~S = \{P_j~|~0 < j < m\}$, where $m (m \leq n)$ is the number of pixels in $S$. 
The set of images and superpixels are denoted by $\I$ and $\S$, respectively. \\

\noindent
\textbf{Region segmentation.}
Regions of interest within images are detached either by parameter-specific methods, \textit{e.g.}, FCN-Net~\cite{Luo2018} and SegNet~\cite{Youssef2018}, or generic methods, \textit{e.g.}, uniform \textit{grids}~\cite{Zahia2018}.
Superpixels are an alternative to both parameter-specific and generic methods, as they split an image $I$ into $k_p$ regions with well-defined borders~\cite{Chino2018}.
Such regions are automatically selected according to a \textit{construction algorithm}. 
Achanta \textit{et al.}~\cite{Achanta2012} conducted a comprehensive survey on superpixel construction algorithms and observed the SLIC method generates high-quality outputs.
We follow their indication and build upon SLIC for performing raw tissue segmentation.\\

\noindent
\textbf{Feature extractor.}
A feature extractor (or \textit{image descriptor}) $\varepsilon$ is a non-bijective function $\varepsilon : \S \rightarrow \mathbb{R}^d$.
Given a superpixel $S \in \S$ of an image $I \in \I$, $\varepsilon$ maps it into a $d$-dimensional feature vector. 
Semantically, the numerical values of the feature vector represent low-level characteristics of the image, such as color, texture, and shape~\cite{Sonka2014}.
Examples of feature extractors applied in wound segmentation include MPEG-7 descriptors Color Layout, Scalable Color, and Color Structure~\cite{Blanco2016}.\\

\noindent
\textbf{Classifier.}
A classifier $c$ is a function $c: \{\mathbb{R}^d, T\} \rightarrow L$, where $L$ is a discrete and disjoint set of labels and $T$ is a training set that conditions the behavior of $c$.
In the context of wounds, $T$ summarizes the set of labeled superpixels $\S' \subseteq \S$, \textit{i.e.}, $T = \{\langle\varepsilon(s_i), l_j\rangle~|~\forall~s_i \in \S'; L = \cup_j l_j\}$ for a feature extractor~$\varepsilon$.
The classifier's \textit{learning algorithm} is a biased method that induces $c$ from $T$ to predict a label $l \in L$ for any superpixel $s_j \in \S$.
Examples of classifiers applied to tissue classification include Na\"ive-Bayes, Multi-Layer Perceptron (MLP), Support Vector Machines (SVM), Instance-based Learning (IbL), and RandomForest~\cite{Mukherjee2014,Kavitha2017,Pereyra2014,Chino2018}.\\

\noindent
\textbf{Convolutional Neural Network}.
A convolutional neural network (CNN) $c_{nn}$ is a function $c_{nn}: \{\mathbb{I}, T\} \rightarrow L$, where $\mathbb{I}$ is the domain of images, $L$ is a discrete and disjoint set of labels, and $T$ is the training set that conditions $c_{nn}$.
In ulcer images, if we set $\mathbb{S} = \S$, then a $c_{nn}$ model can be seen as a classifier that \textit{bypasses} feature extraction, \textit{i.e.}, it uses only the structured set of pixels within superpixels.
While CNNs may present distinct internal topologies, their \textit{learning algorithm} is either directly end-to-end trained from raw pixels of labeled images, or adjusted by a function that performs transfer-learning from a third-party CNN~\cite{Deserno2018,Russakovsky2015}.\\

\noindent
\textbf{Deep-Learning Models}.
A deep-learning (DL) model is the package that includes the CNN topology $c_{nn}$, the CNN learning algorithm $t(c_{nn})$, and the set of labeled and conditioning examples $T$.
DL models have recently surpassed human performance in image classification from basic to complex tasks~\cite{Yuexiang2018,Shin2016}.
A variation of DL models is using a CNN method only for feature extraction, as in the proposal of Nejati \textit{et al.}~\cite{Nejati2018}.
Their approach uniformly divides an image into \textit{patches} that are fed to five convolutional layers.
Three fully-connected layers generate patch representations to be further labeled by an SVM.
The strategy of Zahia \textit{et al.}~\cite{Zahia2018} relies on a similar approach but uses a CNN with nine layers for the labeling of grid regions.
The main drawback with those approaches is they are tightly coupled to specific DL models so that the strategies may not benefit from isolated enhancements on segmentation, feature extraction, or classification.\\

\noindent
\textbf{ImageNet and Transfer Learning.} 
The finding of a suitable DL model for tissue classification requires a massive number of labeled examples, which, in practice, may be either burdensome for experts~\cite{Deserno2018}.
ImageNet Challenge~\cite{Russakovsky2015} is usually employed as the baseline for the definition of new DL models and contains more than $14$ million diversified and labeled images.
Such a baseline enables the topology of CNNs to be adjusted for wound analysis through transfer-learning~\cite{Esteva2017}.
End-to-end trained CNN topologies that reached outstanding results on ImageNet include VGG16~\cite{Simonyan2014}, InceptionV3~\cite{Szegedy2016} and ResNet~\cite{He2016}.\\

\noindent
\textbf{VGG16, InceptionV3, and ResNet.} 
VGG16~\cite{Simonyan2014} is a baseline CNN with 16 convolutional layers that extends AlexNet~\cite{Krizhevsky2012} and enables the handling of image patches, which is suitable for the learning of features from wound images~\cite{Zahia2018,Wang2015}.
On the other hand, InceptionV3~\cite{Szegedy2016} extends GoogLeNet~\cite{Szegedy2015} and provides modules executed in parallel to represent parts of the CNN topology.
The network architecture not only outperformed previous approaches, such as GoogLeNet and VGG16, in the labeling of ImageNet images but also reduced the learning effort in up to $12$ times in comparison to the same competitors.
Finally, ResNet~\cite{He2016} is a recent $152$-layers deep topology that employs residual blocks to guide the learning algorithm on convolutional layers.
A ResNet convolutional neural network was able to solve ImageNet within a 3.6\% error ratio, which poses this CNN on the same tier of VGG16 and InceptionV3~\cite{Russakovsky2015}.\\

\noindent
\textbf{Unbalanced classes.}
A frequent scenario in wound analysis is the uneven distribution of tissue patterns~\cite{Yap2018,Kawahara2016}.
For instance, Nejati \textit{et al.}~\cite{Nejati2018} evaluated a dataset of 350 wound images divided into labeled patches, where most of the labels were related to only three of seven possible classes.
The authors addressed such imbalance by using an approximation of the NP-hard problem that divides the instances into training and test sets.
Likewise, Zahia \textit{et al.}~\cite{Zahia2018} investigated 22 pressure ulcer images, which were divided into 270,762 regions of granulation, 80,636 parts of slough, and 37,146 of necrosis.
The authors used weighted classification metrics for reporting the results. 
Studies on melanoma~\cite{Kawahara2016,Esteva2017} also suggest the use of a sequence of rotations and flips as data augmentation for softening the problem of unbalanced classes.

\section{Materials and Methods}\label{sec:method}

Studies on dermatological wounds rely on small datasets, which hinders the generalization of DL models.
Our method overcomes this drawback by using a divide-and-conquer strategy in which CNNs are requested to handle superpixels instead of single images.
Moreover, we designed the approach in a modular fashion where every stage of the method (segmentation, extraction, and classification) can be set as an external parameter.
Therefore, our method also benefits from enhancements on underlying parameters, such as superpixel construction algorithms and coupled deep-learning models.\\

\noindent
\textbf{Data source.} 
We consider data source~\dataset~\cite{Pereyra2014,Dorileo2010} in the design and evaluation of our proposal.
The set contains $217$ photographies of arterial and venous ulcers in lower limbs with distinct sizes and different healing stages regarding patients with varying skin colors, age, and treatments.
Images represent consecutive evaluations of subjects at the Neurovascular Ulcers Outpatient Clinic of  HCFMRP/USP.
During acquisition white and blue cloths were used to emphasize the contrast between the background and the patients' skin, whereas color patches and rulers were included in the images to facilitate color calibration and normalization~\cite{Dorileo2010}.
All photos were taken with the same camera (Canon EOS 5D0, 2MP, 50mm  macro lens with a polarization filter), angle and distance.
The typical size of an \dataset image is $1747 \times 1165$ pixels with 24 bits-depth.

Experts of HCFMRP/USP were asked to label superpixels from $40$ (out of $217$) ulcered images of distinct and non-related patients within \dataset\footnote{Labeled data is available at: \url{github.com/gu-blanco/ulcer_set}.} by following the four-color class model described in~\cite{Pereyra2014}, which includes labels $L = \{$granulation, fibrin, necrosis, not wound$\}$.
Images were picked by specialists at random, aiming at maximizing diversity, \textit{e.g.}, tissue dominance, skin color, age, and treatment.
The size of superpixels was set to $550$ pixels according to the recommendations in~\cite{Blanco2016,Chino2018}.
As a result, 44,893 superpixels were detached and labeled as follows:
\textit{(i)}~37,187 superpixels with predominant healthy skin area, 
\textit{(ii)}~3,974 fibrin superpixels,
\textit{(iii)}~3,284 superpixels with predominant granulation tissue, and
\textit{(iv)}~448 superpixels of necrosed tissue.
Such instances were employed as the ground-truth for all subsequent evaluations.\\

\begin{figure}[!t]
\centering
\includegraphics[scale=.67]{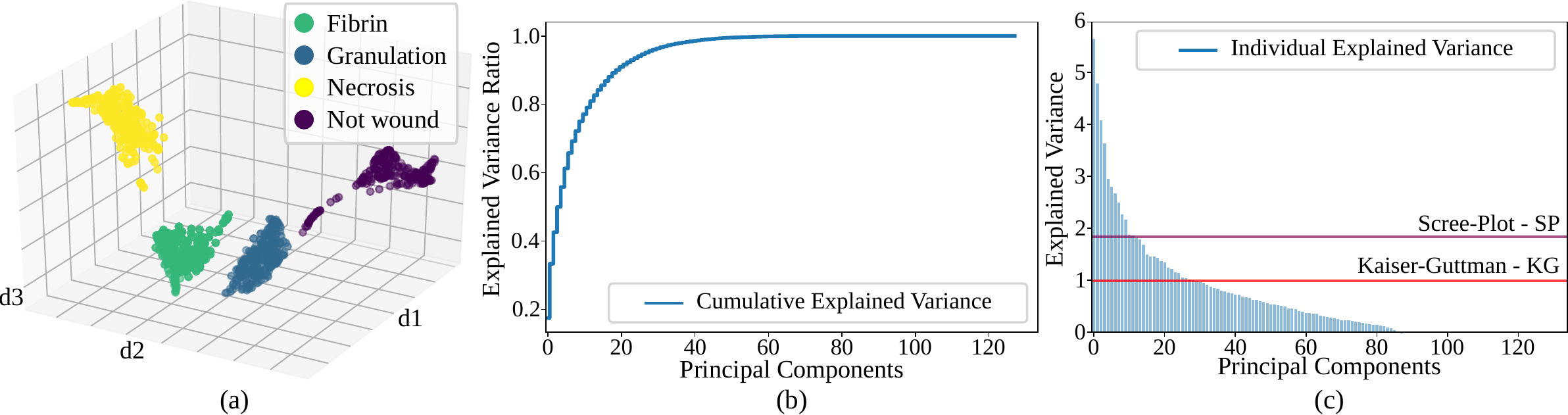}
\caption{Visual analysis of a balanced sample of \dataset superpixels. 
(a)~T-SNE visualization of Color Structure features. 
(b)~PCA Scree Plot.
(c)~PCA individual explained variance plot.}
\label{fig:dataset}
\end{figure}

\noindent
\textbf{Previous approaches.}
Features were extracted from superpixels by Color Layout, Color Structure, and Scalable Color MPEG-7 extractors, which generated multidimensional representations of superpixels in $12$, $128$, and $256$-dimensional spaces, respectively.
Such an extraction serves as data preparation for the visualization of the superpixels' space and also enables the execution of three-stage segmentation pipelines of related studies.
We applied method T-Distributed Neighbor Embedding (T-SNE)~\cite{Van2014} for the visualization of relationships between the superpixels and their labels.
Figure~\ref{fig:dataset}(a) shows the T-SNE visualization of Color Structure vectors after the sampling of the original set of $44,893$ superpixels into $286$ instances per class and by the parameterization of $1,000$ iterations and $200$-step patience.
T-SNE shows the MPEG-7 extractor provides a fair representation of superpixels regarding wounded tissue segmentation since few intersections between instances of different labels were found.

Thus, we also performed a dimensionality reduction before the training of classifiers.
Reductions are suitable for norm-based classifiers~\cite{Mello2018}, which concentrate in medium-to-high dimensional spaces, as the cases of $128$ and $256$-dimensional representations, respectively.
A Principal Components Analysis (PCA) transformation was applied upon such medium-to-high dimensional vectors so that the contribution of each principal component was evaluated regarding explained variances.
Figures~\ref{fig:dataset}(b) and (c) show the behavior of cumulative and individual explained variances for MPEG-7 Color Structure vectors, respectively.
Two distinct criteria were employed to avoid high-dimensional influence and for the finding of the number of reduced dimensions: 
\textit{(i)}~\underline{Kaiser-Guttman (KG)}, which selects components whose individual variances are higher than 1 unit, and
\textit{(ii)}~\underline{Scree-Plot (SP)}, which selects components whose accumulated variance corresponds to the three last quartiles of all variances~\cite{Peres2005}.
KG criterion indicated $31$ components out of $128$ dimensions are required for Color Structure vectors, whereas Scree-Plot criterion selected $12$ components that represent $80.2\%$ of total Color Structure variances.
Analogously, the KG criterion indicated $18$ relevant components out of $256$ dimensions are required for Scalable Color, while Scree-Plot criterion used $6$ components that represent $80.4\%$ of the total Scalable Color variances.\\

\begin{table}[!t]
\centering
\caption{MPEG-7 extraction and classification of superpixels regarding AUC and Cohen-Kappa Coefficient (CKC).}\label{tab:classification}
\begin{tabular}{p{1.23cm}|p{1.41cm}|p{1.41cm}|p{1.41cm}|p{1.41cm}|p{1.41cm}|p{1.41cm}|p{1.41cm}|p{1.41cm}}\hline \hline

& 
\footnotesize \textbf{Random\-Forest} & 
\footnotesize \textbf{Random\-Tree} & 
\footnotesize \textbf{SVM} & 
\footnotesize \textbf{MLP} & 
\footnotesize \textbf{Na\"{i}ve\--Bayes} & 
\footnotesize \textbf{Bayes\--Net} & 
\footnotesize \textbf{IbL-$L_2$} & 
\footnotesize\textbf{IbL-$L_1$} \\ \hline \hline
                     
\rowcolor[HTML]{CBCEFB} 
\footnotesize Color$\> \> \> \> \> \> \> \>  \> \> \> \>  $ Layout           & 
\footnotesize 0.902 AUC 0.448 CKC    & 
\footnotesize 0.698 AUC 0.374 CKC    & 
\footnotesize 0.646 AUC 0.331 CKC    & 
\footnotesize 0.879 AUC 0.427 CKC    & 
\footnotesize 0.822 AUC 0.336 CKC    & 
\footnotesize 0.853 AUC 0.376 CKC    & 
\footnotesize 0.696 AUC 0.376 CKC    & 
\footnotesize 0.397 AUC 0.376 CKC       \\ \hline

\rowcolor[HTML]{9AFF99} 
\footnotesize Color Structure      & 
\footnotesize 0.952 AUC 0.665 CKC             & 
\footnotesize \underline{0.791 AUC} 0.548 CKC & 
\footnotesize \underline{0.675 AUC} 0.410 CKC    & 
\footnotesize 0.880 AUC 0.563 CKC    & 
\footnotesize \underline{0.815 AUC} 0.149 CKC            & 
\footnotesize \underline{0.875 AUC} 0.438 CKC          & 
\footnotesize 0.815 AUC 0.592 CKC       & 
\footnotesize 0.812 AUC 0.583 CKC       \\

\rowcolor[HTML]{8CD18C} 
\footnotesize Color Structure KG & 
\footnotesize 0.947 AUC 0.641 CKC    & 
\footnotesize 0.790 AUC 0.542 CKC            & 
\footnotesize 0.630 AUC 0.331 CKC    & 
\footnotesize \underline{0.921 AUC} 0.568 CKC    & 
\footnotesize 0.704 AUC 0.140 CKC            & 
\footnotesize 0.872 AUC 0.351 CKC          & 
\footnotesize 0.806 AUC 0.577 CKC       & 
\footnotesize 0.809 AUC 0.582 CKC       \\

\rowcolor[HTML]{709E6F} 
\footnotesize Color Structure SP & 
\footnotesize 0.940 AUC 0.604 CKC             & 
\footnotesize 0.777 AUC 0.516 CKC            & 
\footnotesize 0.604 AUC 0.262 CKC    & 
\footnotesize 0.900 AUC 0.531 CKC    & 
\footnotesize 0.706 AUC 0.133 CKC            & 
\footnotesize 0.850 AUC 0.333 CKC          & 
\footnotesize 0.794 AUC 0.551 CKC       &
\footnotesize 0.794 AUC 0.549 CKC       \\ \hline

\rowcolor[HTML]{FFCE93} 
\footnotesize Scalable Color       & 
\footnotesize 0.947 AUC 0.606 CKC             & 
\footnotesize 0.760 AUC 0.485 CKC            & 
\footnotesize 0.601 AUC 0.207 CKC    & 
\footnotesize 0.815 AUC 0.381 CKC    & 
\footnotesize 0.502 AUC 0.000 CKC            & 
\footnotesize 0.619 AUC 0.058 CKC          & 
\footnotesize \underline{0.827 AUC} 0.613 CKC       & 
\footnotesize 0.796 AUC 0.551 CKC       \\

\rowcolor[HTML]{F4BE7E} 
\footnotesize Scalable Color   KG  & 
\footnotesize \underline{0.955 AUC} 0.645 CKC    & 
\footnotesize 0.782 AUC 0.534 CKC            & 
\footnotesize 0.600 AUC 0.200 CKC    & 
\footnotesize 0.806 AUC 0.330 CKC    & 
\footnotesize 0.497 AUC 0.000 CKC            & 
\footnotesize 0.650 AUC 0.000 CKC          & 
\footnotesize 0.807 AUC 0.569 CKC       & 
\footnotesize \underline{0.813 AUC} 0.584 CKC       \\

\rowcolor[HTML]{D6A05F} 
\footnotesize Scalable Color   SP  & 
\footnotesize 0.906 AUC 0.507 CKC             & 
\footnotesize 0.715 AUC 0.414 CKC            & 
\footnotesize 0.600 AUC 0.200 CKC    & 
\footnotesize 0.807 AUC 0.310 CKC    & 
\footnotesize 0.495 AUC 0.000 CKC            & 
\footnotesize 0.774 AUC 0.243 CKC          & 
\footnotesize 0.740 AUC 0.439 CKC       & 
\footnotesize 0.746 AUC 0.452 CKC   \\ \hline \hline   
\end{tabular}
\end{table}

\noindent
\textbf{Machine-learning classification.} A representative set of classifiers was trained with seven multidimensional superpixel representations, namely 
Color Layout,
Color Structure,
Color Structure -- KG,
Color Structure -- SP,
Scalable Color,
Scalable Color -- KG, and
Scalable Color -- SP.
We examined RandomTree, Na\"{i}ve-Bayes e Bayes-Net classifiers, as well as methods RandomForest, SVM, MLP, and IbL, employed in previous studies for superpixel labeling.
A broad set of parameters was tested for the fine-tuning of the classifiers and results indicated
\textit{(i)}~$10$ trees and Gini as the objective function for RandomForest, 
\textit{(ii)}~$02$ fully-connected hidden layers with r-prop learning algorithm for MLP,
\textit{(iii)}~$k=1$ neighbor for IbL regarding both $L_1$ and $L_2$ norms, and 
\textit{(iv)}~Hill-Climbing heuristic for Bayes-Net construction.
Other parameters were set with the default values found in the Weka \textit{framework}\footnote{Available at: \url{www.cs.waikato.ac.nz/ml/weka/}}.
The classifiers and extractors ($56$ combinations) were evaluated by a $10$-fold cross-validation procedure on $40$ labeled images.
Table~\ref{tab:classification} shows the results regarding Area under the ROC Curve (AUC) and Cohen-Kappa Coefficient (CKC).
Color Structure and Scalable Color provided more suitable representations than Color Layout, whereas dimensionality reduction produced three of the eight best scenarios -- see underline values in Table~\ref{tab:classification}.
Although MLP and Bayes-Net outperformed most of the competitors regarding Color Layout and Color Structure representations, they underperformed for Scalable Color vectors.
RandomForest outperformed every competitor and reached the highest AUC score through the labeling of Scalable Color -- KG representation (0.955).\\

\noindent
\textbf{\system approach.} Our proposal, named \system\footnote{\system is available at: \url{github.com/gu-blanco/qtdu/}}, \textit{bypasses} feature extraction by relying on CNNs for finding the most suitable representations of wounded tissues within dermatological ulcers.
Therefore, \system main parameters are related to the underlying DL model used for the superpixel labeling task, \textit{i.e.}, CNN topology, its learning algorithm, and the conditioning training set.
In particular, \system builds upon results of ImageNet~\cite{Russakovsky2015} and the DL models in~\cite{Esteva2017} and \cite{Han2018} for defining two CNN candidates: ResNet and InceptionV3.
In the evaluations, we used the same learning parameters of those previous studies, \textit{i.e.}, learning ratio of $0.001$, momentum to $0.88$, training and decay patience of $200$ (the learning patient of $50$ epochs), loss function as `cat\-e\-gorical cross entropy', and batch size of $24$.

\begin{figure}[!t]
\centering
\includegraphics[scale=1]{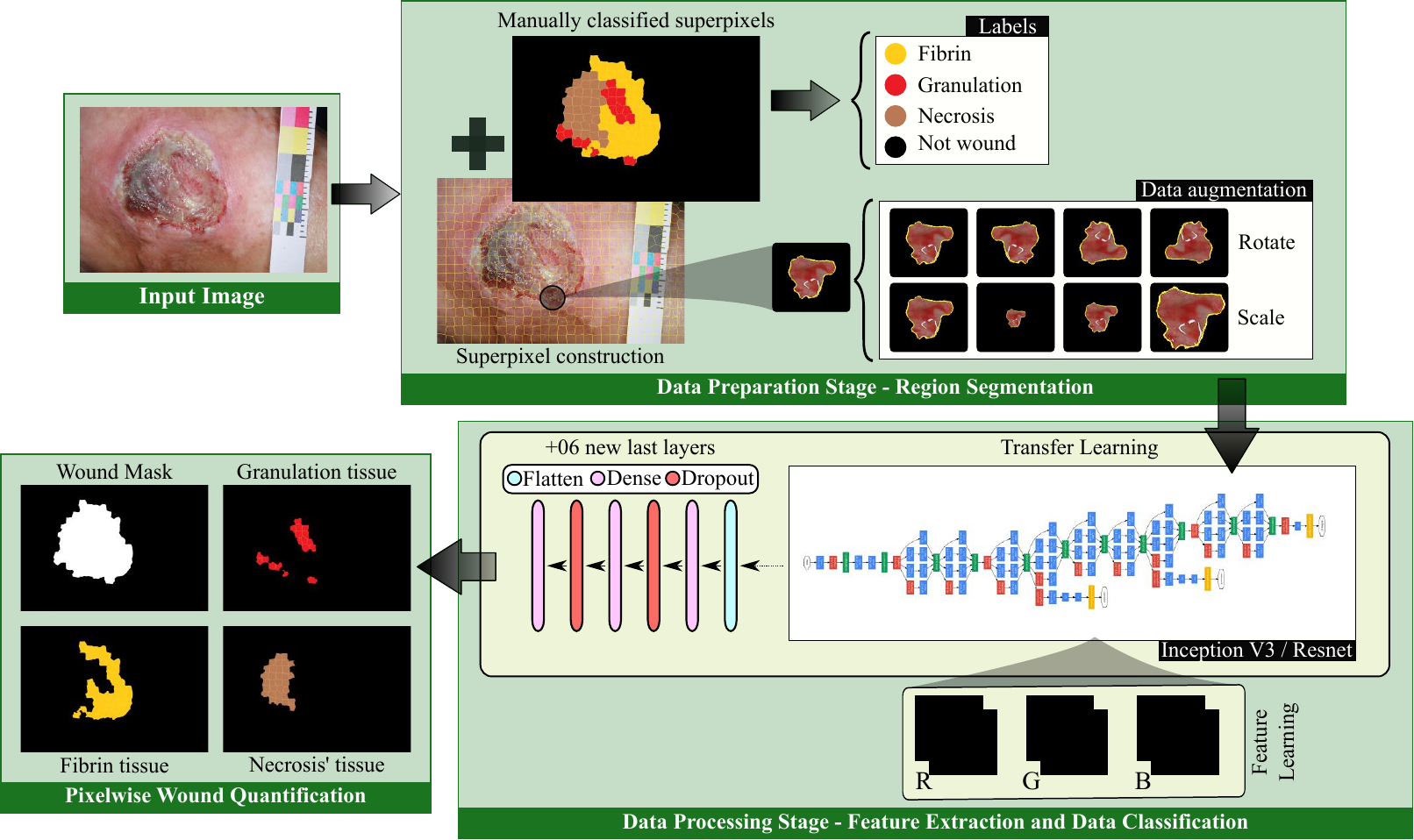}
\caption{\system overall architecture. 
In the learning phase, the image is divided into superpixels that adjust the underlying CNN and the new six layers.
The final ``mask'' is obtained by joining superpixels labeled as `Not Wound'.}
\label{fig:approach}
\end{figure}

Additionally, we prepared the underlying CNN to handle overfitting since both ResNet and InceptionV3 include millions of adjustable parameters.
Alternatives for such task include the use of layer regularization and the insertion of several dropout tiers, which substantially modify the network topology.
However, our approach requires the architecture to be as similar to the underlying topology as possible, so that a more comprehensive transfer-learning can be performed from third-party CNNs.
The management of overfitting in such a context relies on
\textit{(i)}~\underline{data augmentation}, applied to superpixels for increasing the number of instances in the training set, and
\textit{(ii)}~\underline{careful addition of six new layers}, included at the end of the underlying CNN.
We followed the regularization hints of Hinton \textit{et al.}~\cite{Hinton2012} for including new levels as three fully-connected tiers interleaved with two dropout layers (\textit{i.e.}, Dense--Dropout--Dense--Dropout--Dense), in which the final layer provides the label.
Activation function ReLU with 512 units (nodes) was set to the first two dense layers, whereas a \textit{softmax} function with four outputs was used in the last layer.
As for dropouts, we applied a 0.5 rating so that half of the activation units may be nullified by the generalization routine.

Figure~\ref{fig:approach} illustrates the three-stage \system pipeline, where the image is first divided into superpixels whose construction enables homogeneous regions to be kept into single blocks (regardless of their shape) according to similarity-based metrics. 
The learning phase consists of training the original CNN weights out of ImageNet and random variables at the new layers, which are end-to-end adjusted according to superpixels.\\

\noindent
\textbf{Data Augmentation.}
The number of \dataset instances was augmented through the use of
\textit{(i)}~rotations of $90^{o}$,
\textit{(ii)}~scale in and out by a 0.2 factor, and
\textit{(iii)}~vertical and horizontal mirroring.
Such a choice of parameters enables the targeting of wounded tissue symmetry (horizontal/vertical flip) and different wound sizes (zoom in).
As a result, 179,572 instances and four labels were used for the training of \system.
Our implementation employs TensorFlow ImageGenerator for data augmentation, which generates augmented superpixels at runtime and avoids loading all instances into main memory in the learning phase.\\

\noindent
\textbf{Pixelwise area quantification.} \system divides unsegmented images into superpixels that are further fused according to their labels into four distinct regions, namely \textit{(i)}~wound mask, 
\textit{(ii)}~granulation, 
\textit{(iii)}~fibrin, and 
\textit{(iv)}~necrotic tissues. 
Such fused regions are quantified regarding the number of pixels per superpixel and the total number of pixels within the image.

\section{Experiments}\label{sec:experiments}

\system was tested on an Ubuntu 16.04.4 LTS OS running on a local cluster with two nodes with 2560 GPU cores at 1607 MHz each, 64 Gb shared RAM.
The evaluations aimed at
\textit{(i)}~determining the most suitable settings for \system, and
\textit{(ii)}~quantifying \system improvements regarding wound segmentation in comparison to previous machine-learning approaches and patch-based CNNs.
Accordingly, we selected the best performances of Table~\ref{tab:classification} and compared them to \system with underlying CNNs InceptionV3\footnote{\url{keras.io/applications/\#inceptionv3}} and ResNet\footnote{\url{keras.io/applications/\#resnet50}}. 
The next four comparisons were performed according to a \textit{$10$-fold cross-validation} procedure.\\

\begin{figure}[!htb]
\centering
\includegraphics[scale=1.03]{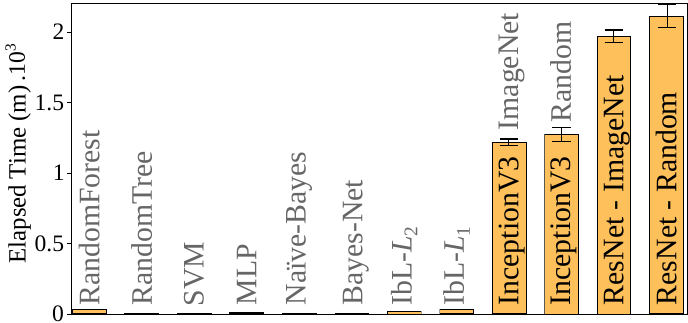}
\caption{Average and standard deviation elapsed time for the training of machine-learning classifiers and \system with and without random weights.}
\label{fig:time}
\end{figure}

\noindent
\textbf{CNN Training.} 
We measured the time spent on the training of two \system underlying CNNs, InceptionV3 and ResNet, with and without initial random weights.
Additionally, we also measured the time spent on the training of every classifier with features indicated by the highest AUCs in Table~\ref{tab:classification}.
Figure~\ref{fig:time} shows the average and standard deviation time for ten runnings of the learning algorithms without on-the-fly data augmentation.
While the quality of CNNs with random and ImageNet pertaining weights was similar, the training with random weights demanded more time, on average, with a greater standard deviation.
In particular, the training of InceptionV3 required 1,356.1$\pm$64.19m and 1.208,6$\pm$36.41m for random and ImageNet weights, respectively.
Analogously, the training of ResNet took 2,118.8$\pm$89.88m and 2,050.6$\pm$35.49m regarding random and ImageNet parameters, respectively.
Results also showed RandomForest was nearly three orders of magnitude faster than the cheapest CNN.
Finally, we measured the time spent on image classification after training the underlying CNNs.
On average, \system with InceptionV3 required 1.41$\pm$0.34s, whereas \system with ResNet took 1.95$\pm$0.19s for labeling a superpixel.
All evaluations were performed on the same GPU-based cluster, Python $3.6.3$, Keras, and Scikit-learn $0.2$.

\begin{table}[!t]
\centering
\caption{\system \textit{vs.} machine-learning-based  comparison regarding the task of labeling superpixels.}\label{tab:results}
\begin{tabular}{p{4.2cm}|c|c|c|c|c} \hline \hline

 & 
\footnotesize \textbf{CKC} & 
\footnotesize \textbf{F1-Score}   & 
\footnotesize \textbf{Sensitivity} & 
\footnotesize \textbf{Specificity} & 
\footnotesize \textbf{AUC}  \\ \hline
 
\footnotesize RandomForest w/ Scalable Color -- KG         & 
\footnotesize $0.645\pm0.017$  & 
\footnotesize $0.897\pm0.018$  & 
\footnotesize $0.904\pm0.016$  & 
\footnotesize $0.896\pm0.016$  & 
\footnotesize $0.955\pm0.016$ \\ \hline
                                   
\footnotesize RandomTree w/ Color Structure      & 
\footnotesize $0.548\pm0.013$  & 
\footnotesize $0.858\pm0.011$  & 
\footnotesize $0.875\pm0.014$  & 
\footnotesize $0.866\pm0.016$  & 
\footnotesize $0.791\pm0.015$ \\ \hline
                                   
\footnotesize SVM w/ Color Structure             & 
\footnotesize $0.410\pm0.015$  & 
\footnotesize $0.876\pm0.019$  & 
\footnotesize $0.591\pm0.015$  & 
\footnotesize $0.865\pm0.021$  & 
\footnotesize $0.675\pm0.017$ \\ \hline
                                   
\footnotesize MLP w/ Color Structure -- KG       & 
\footnotesize $0.568\pm0.010$  & 
\footnotesize $0.819\pm0.092$  & 
\footnotesize $0.821\pm0.104$  & 
\footnotesize $0.827\pm0.080$  & 
\footnotesize $0.921\pm0.011$ \\ \hline
                                   
\footnotesize Na\"{i}ve-Bayes w/ Color Structure & 
\footnotesize $0.149\pm0.004$  & 
\footnotesize $0.408\pm0.004$  & 
\footnotesize $0.866\pm0.003$  & 
\footnotesize $0.357\pm0.003$  & 
\footnotesize $0.815\pm0.004$ \\ \hline
                                   
\footnotesize Bayes-Net w/ Color Structure       & 
\footnotesize $0.438\pm0.013$  & 
\footnotesize $0.821\pm0.019$  & 
\footnotesize $0.856\pm0.019$  & 
\footnotesize $0.792\pm0.027$  & 
\footnotesize $0.875\pm0.021$ \\ \hline
                                   
\footnotesize IbL-$L_2$ w/ Scalable Color        & 
\footnotesize $0.613\pm0.019$  & 
\footnotesize $0.892\pm0.016$  & 
\footnotesize $0.879\pm0.026$  &
\footnotesize $0.870\pm0.023$  & 
\footnotesize $0.827\pm0.021$ \\ \hline
                                   
\footnotesize IbL-$L_1$ w/ Scalable Color -- KG  & 
\footnotesize $0.584\pm0.023$  & 
\footnotesize $0.851\pm0.020$  & 
\footnotesize $0.856\pm0.037$  & 
\footnotesize $0.853\pm0.036$  & 
\footnotesize $0.813\pm0.034$ \\ \hline \hline
                 
% \footnotesize \textbf{\system -- VGG}                  &
% \footnotesize \textbf{0.711$\pm$0.002}  & 
% \footnotesize \textbf{0.964$\pm$0.002} & 
% \footnotesize \textbf{0.963$\pm$0.004}  & 
% \footnotesize \textbf{0.967$\pm$0.013} & 
% \footnotesize \textbf{0.982$\pm$0.007} \\ \hline
                                   
\footnotesize \textbf{\system w/ InceptionV3}            & 
\footnotesize \textbf{0.716$\pm$0.001}  & 
\footnotesize \textbf{0.969$\pm$0.007}  & 
\footnotesize \textbf{0.968$\pm$0.006}  & 
\footnotesize \textbf{0.971$\pm$0.012} & 
\footnotesize \textbf{0.986$\pm$0.018} \\ \hline

\footnotesize \textbf{\system w/ ResNet}                  &
\footnotesize \textbf{0.721$\pm$0.001}  & 
\footnotesize \textbf{0.971$\pm$0.004} & 
\footnotesize \textbf{0.970$\pm$0.004}  & 
\footnotesize \textbf{0.974$\pm$0.007} & 
\footnotesize \textbf{0.986$\pm$0.012} \\ \hline\hline
\end{tabular}
\end{table}

\noindent
\textbf{\system \textit{vs.} Machine-learning Classification.}
Table~\ref{tab:results} shows an overall comparison between machine-learning-based classification and \system with distinct parameterizations regarding Cohen-Kappa Coefficient, F1-Score, Sensitivity, Specificity, and AUC.
\system outperformed machine-learning classification in every scenario and metric.
In particular, \system with InceptionV3 outperformed the classifier (RandomForest) in up to $6.4\%$ and $7.6\%$ regarding Sensitivity and Specificity, respectively.
Likewise, \system with ResNet outperformed RandomForest in up to $7.3\%$ and $8.7\%$ regarding Sensitivity and Specificity, respectively.
\system with ResNet also outperformed the best machine-learning-based approach in up to $3.6\%$ regarding AUC and reached substantially higher values of Cohen-Kappa Coefficient and F1-Score in all comparisons.

\begin{figure}[!htb]
\centering
\includegraphics[scale=.533]{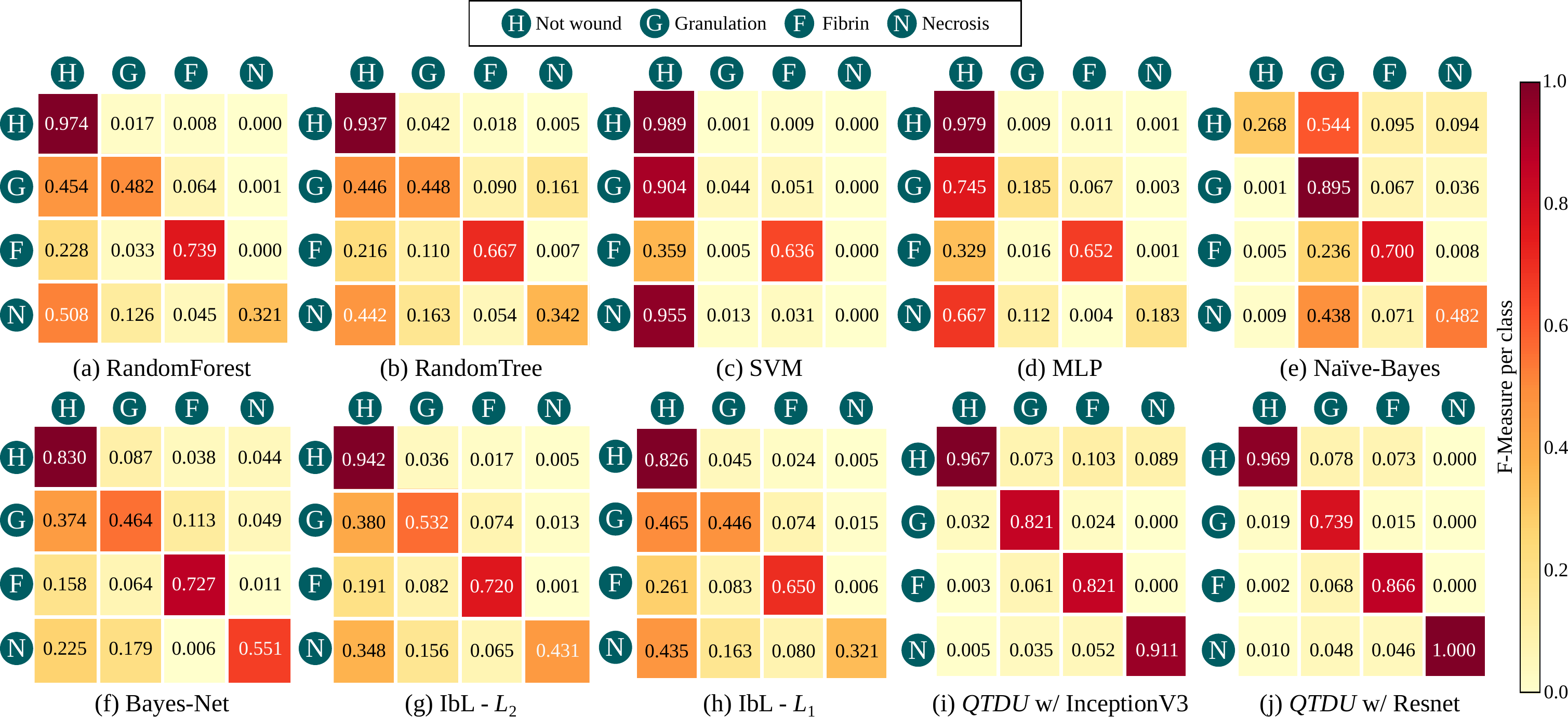}
\caption{F1-Scores reached by different superpixel classes according to both \system and machine-learning methods. 
The scale of values ranges in the [0--1] interval.
The closer to $1$, the better the score.}
\label{fig:heatmap}
\end{figure}

Figure~\ref{fig:heatmap} provides a comparison of machine-learning approaches and \system regarding F1-Scores per class.
Results indicate machine-learning methods achieved measures higher than 0.7 for only two of four classes at the same time.
On the other hand, confusion matrices of Figures~\ref{fig:heatmap}(i -- j) show \system provided a more uniform result per tissue with the lowest F1-Score of 0.739.
Finally, Figures~\ref{fig:heatmap}(a -- h) show the trade-offs between machine-learning approaches, which are more effective for particular tissues depending on the inducing bias and the feature extractor.
For instance, RandomForest was slightly better than \system for the detection of ``not wound'' superpixels, a scenario in which our approach delivered false positives at a very low ratio.
Last, but not least, \system with ResNet achieved results slightly better than InceptionV3 for the labeling of \textit{not wound}, \textit{fibrin}, and \textit{necrosis} superpixels.
Aimed at investigating the significance of such differences, we applied a hypothesis test on \system and machine-learning solutions.\\

\noindent
\textbf{Ranking test.}
In this evaluation, we applied a \textit{leave-one-out} procedure for the labeling of the $40$ uncorrelated images in our ground-truth, instead of using average values from a $10$-fold cross-validation routine.
Accordingly, a distinct Cohen-Kappa Coefficient per image was assigned to machine-learning-based methods and \system.
Next, we performed a hypothesis test to evaluate whether significant differences existed between the coefficients achieved by the competitors. 
In particular, we applied the well-known Friedman ranking test~\cite{Demsar2006} to assess such differences.
By using a significance level of 0.01, we obtained a $p$-value of $7.54 \cdot 10^{-20}$ and, consequently, we rejected the Friedman null hypothesis that differences are due to random sampling, and concluded at least one of the performances differs from the others.

After the rejection of Friedman's null hypothesis, we applied the Nemenyi \textit{post-test}~\cite{Garcia2008} for comparing pairs of approaches according to confidence intervals of 99\%, 95\%, and 90\%.
Table~\ref{fig:hypothesis_test} shows the heatmap based on the Nemenyi $p$-values for each pair of compared approaches (lines \textit{vs.} columns). 
\system outperformed every machine-learning competitor within strong significant levels, whereas some machine-learning methods, \textit{e.g.}, RandomForest, also outperformed others within significant margins.
Moreover, although no significant difference was observed between \system with either InceptionV3 or ResNet, results indicate ResNet outperformed other approaches with slightly better $p$-values in comparison to InceptionV3.\\

\begin{table}[!t]
\centering
\caption{Heatmap of $p$-values regarding the pairwise comparison of machine-learning methods and \system in the labeling of dermatological wound photos (methods are compared as lines \textit{vs.} columns).}
\label{fig:hypothesis_test}
\small
\begin{tabular}{p{1.1cm}|p{1cm}|p{1.1cm}|p{1.08cm}|p{1.08cm}|p{1.08cm}|p{1.08cm}|p{1.08cm}|p{1.08cm}|p{0.9cm}|p{0.9cm}} \hline \hline              
& \scriptsize{\textbf{Random Forest}}
& \scriptsize{\textbf{Random Tree}} 
& \scriptsize \textbf{SVM} 
& \scriptsize \textbf{MLP}   
& \scriptsize{\textbf{\begin{tabular}[c]{@{}l@{}}Na\"{i}ve-\\ Bayes\end{tabular}}}
& \scriptsize{\textbf{\begin{tabular}[c]{@{}l@{}}Bayes-\\ Net\end{tabular}}}
& \scriptsize{\textbf{\begin{tabular}[c]{@{}l@{}}IbL\\ $L_2$\end{tabular}}}  
& \scriptsize{\textbf{\begin{tabular}[c]{@{}l@{}}IbL\\ $L_1$\end{tabular}}}  
& \scriptsize{\textbf{\begin{tabular}[c]{@{}l@{}}\system-\\ Inc.V3\end{tabular}}}
& \scriptsize{\textbf{\begin{tabular}[c]{@{}l@{}}\system-\\ ResNet\end{tabular}}} \\ \hline 

\scriptsize{\textbf{Random Forest}} 
& \scriptsize -                                                             
& \cellcolor[HTML]{DC13E8}\scriptsize 1 $\cdot 10^{-13}$
& \cellcolor[HTML]{DC13E8} \scriptsize 0           
& \cellcolor[HTML]{DC13E8}\scriptsize 5 $\cdot 10^{-11}$ 
& \cellcolor[HTML]{DC13E8} \scriptsize 0                              
& \cellcolor[HTML]{DC13E8} \scriptsize 0
& \cellcolor[HTML]{D258D9}\scriptsize 9 $\cdot 10^{-2}$  
& \cellcolor[HTML]{DC13E8}\scriptsize 3 $\cdot 10^{-6}$  
& \cellcolor[HTML]{FBBAFF} \scriptsize 1
& \cellcolor[HTML]{FBBAFF} \scriptsize 1   \\ \hline
   
\scriptsize{\textbf{Random Tree}}   
& \cellcolor[HTML]{FBBAFF} \scriptsize 1
& \scriptsize -                                                             
& \cellcolor[HTML]{DC13E8} \scriptsize 3  $\cdot 10^{-6}$  
& \cellcolor[HTML]{FBBAFF} \scriptsize 1           
& \cellcolor[HTML]{DC13E8}\scriptsize 2 $\cdot 10^{-13}$               
& \cellcolor[HTML]{B047B6}\scriptsize 2 $\cdot 10^{-2}$
& \cellcolor[HTML]{FBBAFF} \scriptsize 1           
& \cellcolor[HTML]{FBBAFF} \scriptsize 1           
& \cellcolor[HTML]{FBBAFF} \scriptsize 1                                    
& \cellcolor[HTML]{FBBAFF} \scriptsize 1   \\ \hline

\scriptsize{\textbf{SVM}}
& \cellcolor[HTML]{FBBAFF} \scriptsize 1                                  
& \cellcolor[HTML]{FBBAFF} \scriptsize 1                                    
& \scriptsize -                                    
& \cellcolor[HTML]{FBBAFF} \scriptsize 1           
& \cellcolor[HTML]{FBBAFF}\scriptsize 2 $\cdot 10^{-1}$
& \cellcolor[HTML]{FBBAFF} \scriptsize 1                                    
& \cellcolor[HTML]{FBBAFF} \scriptsize 1           
& \cellcolor[HTML]{FBBAFF} \scriptsize 1           
& \cellcolor[HTML]{FBBAFF} \scriptsize 1  
& \cellcolor[HTML]{FBBAFF} \scriptsize 1        \\ \hline

\scriptsize{\textbf{MLP}}
& \cellcolor[HTML]{FBBAFF} \scriptsize 1                                      
& \cellcolor[HTML]{FBBAFF}\scriptsize 6 $\cdot 10^{-1}$                   
& \cellcolor[HTML]{DC13E8}\scriptsize 9 $\cdot 10^{-12}$ 
& \scriptsize -                          
& \cellcolor[HTML]{DC13E8}\scriptsize 1 $\cdot 10^{-13}$                 
& \cellcolor[HTML]{DC13E8}\scriptsize 3 $\cdot 10^{-6}$                   
& \cellcolor[HTML]{FBBAFF} \scriptsize 1           
& \cellcolor[HTML]{FBBAFF} \scriptsize 1           
& \cellcolor[HTML]{FBBAFF} \scriptsize 1                                       
& \cellcolor[HTML]{FBBAFF} \scriptsize 1 \\ \hline

\scriptsize{\textbf{\begin{tabular}[c]{@{}l@{}}Na\"{i}ve-\\ Bayes\end{tabular}}  } 
& \cellcolor[HTML]{FBBAFF} \scriptsize 1                     
& \cellcolor[HTML]{FBBAFF} \scriptsize 1             
& \cellcolor[HTML]{FBBAFF} \scriptsize 1      
& \cellcolor[HTML]{FBBAFF} \scriptsize 1     
& \scriptsize -                                                                
& \cellcolor[HTML]{FBBAFF} \scriptsize 1         
& \cellcolor[HTML]{FBBAFF} \scriptsize 1   
& \cellcolor[HTML]{FBBAFF} \scriptsize 1   
& \cellcolor[HTML]{FBBAFF} \scriptsize 1                                       
& \cellcolor[HTML]{FBBAFF} \scriptsize 1  \\ \hline

\scriptsize{\textbf{\begin{tabular}[c]{@{}l@{}}Bayes-\\ Net\end{tabular}}     }
& \cellcolor[HTML]{FBBAFF}\scriptsize 1     
& \cellcolor[HTML]{FBBAFF}\scriptsize 1       
& \cellcolor[HTML]{FBBAFF}\scriptsize 6 $\cdot 10^{-1}$  
& \cellcolor[HTML]{FBBAFF}\scriptsize 1     
& \cellcolor[HTML]{DC13E8}\scriptsize 1 $\cdot 10^{-4}$
& \scriptsize -                                   
& \cellcolor[HTML]{FBBAFF}\scriptsize 1    
& \cellcolor[HTML]{FBBAFF}\scriptsize 1   
& \cellcolor[HTML]{FBBAFF}\scriptsize 1   
& \cellcolor[HTML]{FBBAFF}\scriptsize 1 \\ \hline

\scriptsize{\textbf{IbL - $L_2$}   }
& \cellcolor[HTML]{FBBAFF} \scriptsize 1
& \cellcolor[HTML]{DC13E8}\scriptsize 4 $\cdot 10^{-8}$
& \cellcolor[HTML]{DC13E8} \scriptsize 0
& \cellcolor[HTML]{DC13E8}\scriptsize 1 $\cdot 10^{-3}$  
& \cellcolor[HTML]{DC13E8} \scriptsize 0                                       
& \cellcolor[HTML]{DC13E8}\scriptsize 8 $\cdot 10^{-14}$
& \scriptsize - 
& \cellcolor[HTML]{FBBAFF}\scriptsize 3 $\cdot 10^{-1}$
& \cellcolor[HTML]{FBBAFF} \scriptsize 1                                       
& \cellcolor[HTML]{FBBAFF} \scriptsize 1 \\ \hline

\scriptsize{\textbf{IbL - $L_1$}    }
& \cellcolor[HTML]{FBBAFF} \scriptsize 1
& \cellcolor[HTML]{DC13E8}\scriptsize 9 $\cdot 10^{-3}$
& \cellcolor[HTML]{DC13E8}\scriptsize 1 $\cdot 10^{-13}$ 
& \cellcolor[HTML]{FBBAFF}\scriptsize 8 $\cdot 10^{-1}$  
& \cellcolor[HTML]{DC13E8} \scriptsize 0
& \cellcolor[HTML]{DC13E8}\scriptsize 5 $\cdot 10^{-11}$
& \cellcolor[HTML]{FBBAFF} \scriptsize 1
& \scriptsize - 
& \cellcolor[HTML]{FBBAFF} \scriptsize 1
& \cellcolor[HTML]{FBBAFF} \scriptsize 1 \\ \hline

\scriptsize{\textbf{\begin{tabular}[c]{@{}l@{}}\system-\\ Inc.V3\end{tabular}} }
& \cellcolor[HTML]{D258D9}\scriptsize 9 $\cdot 10^{-2}$
& \cellcolor[HTML]{DC13E8} \scriptsize 0                                       
& \cellcolor[HTML]{DC13E8} \scriptsize 0
& \cellcolor[HTML]{DC13E8}\scriptsize 1 $\cdot 10^{-13}$ 
& \cellcolor[HTML]{DC13E8} \scriptsize 0                                       
& \cellcolor[HTML]{DC13E8} \scriptsize 0
& \cellcolor[HTML]{DC13E8}\scriptsize 2 $\cdot 10^{-7}$  
& \cellcolor[HTML]{DC13E8}\scriptsize 1 $\cdot 10^{-13}$ 
& \scriptsize - 
& \cellcolor[HTML]{FBBAFF} \scriptsize 1 \\ \hline

\scriptsize{\textbf{\begin{tabular}[c]{@{}l@{}}\system-\\ ResNet\end{tabular}}}
& \cellcolor[HTML]{DC13E8}\scriptsize 4 $\cdot 10^{-4}$
& \cellcolor[HTML]{DC13E8} \scriptsize 0                                       
& \cellcolor[HTML]{DC13E8} \scriptsize 0
& \cellcolor[HTML]{DC13E8} \scriptsize 0
& \cellcolor[HTML]{DC13E8} \scriptsize 0                                       
& \cellcolor[HTML]{DC13E8} \scriptsize 0
& \cellcolor[HTML]{DC13E8}\scriptsize 9 $\cdot 10^{-12}$
& \cellcolor[HTML]{DC13E8}\scriptsize 1 $\cdot 10^{-13}$ 
& \cellcolor[HTML]{FBBAFF}\scriptsize 9 $\cdot 10^{-1}$
& \scriptsize -  \\ \hline \hline
\end{tabular}
\begin{tabular}{
>{\columncolor[HTML]{DC13E8}}l 
>{\columncolor[HTML]{DC13E8}}l 
>{\columncolor[HTML]{B047B6}}l 
>{\columncolor[HTML]{B047B6}}l 
>{\columncolor[HTML]{D258D9}}l 
>{\columncolor[HTML]{D258D9}}l 
>{\columncolor[HTML]{FBBAFF}}l 
>{\columncolor[HTML]{FBBAFF}}l 
>{\columncolor[HTML]{FBBAFF}}l } \hline
\multicolumn{2}{|l|}{\cellcolor[HTML]{DC13E8}  \scriptsize \textbf{99\% confidence}} 
& \multicolumn{2}{l|}{\cellcolor[HTML]{B047B6} \scriptsize \textbf{95-99\% confidence}} 
& \multicolumn{2}{l|}{\cellcolor[HTML]{D258D9} \scriptsize \textbf{90-95\% confidence}} 
& \multicolumn{3}{l|}{\cellcolor[HTML]{FBBAFF} \scriptsize \textbf{No significant differences}}\\ \hline
\end{tabular}
\normalsize
\end{table}

\begin{figure}[!htb]
\centering
\includegraphics[scale=1.03]{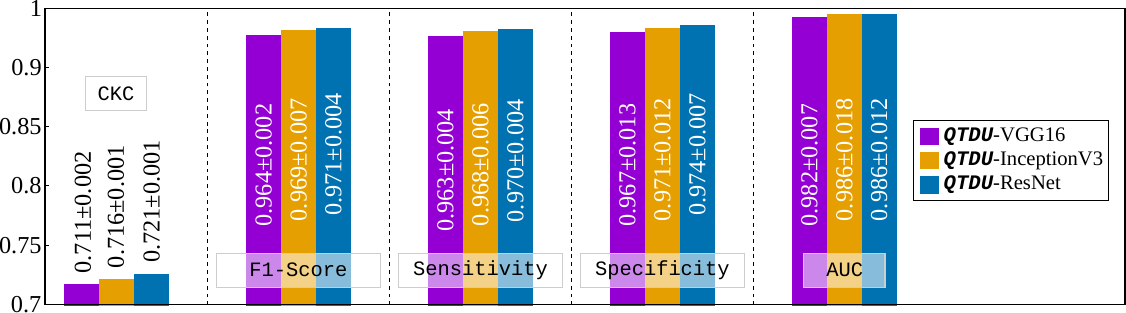}
\caption{
\system with underlying CNNs VGG16, InceptionV3, and ResNet comparison regarding Cohen-Kappa Coefficient (CKC), F1-Score, Sensitivity, Specificity, and AUC.}
\label{fig:vsVGG}
\end{figure}

\noindent
\textbf{VGG16, InceptionV3, and ResNet.}
Although state-of-the-art CNNs InceptionV3 and ResNet were employed as \system underlying parameters in previous evaluations, patch-based VGG16 can be seamlessly used as well.
Figure~\ref{fig:vsVGG} shows the comparison between VGG16, InceptionV3, and ResNet as \system underlying CNNs regarding the mean and standard deviation of Cohen-Kappa Coefficient, F1-Score, Sensitivity, Specificity, and AUC gathered from the same experimental testbed of evaluation ``\system \textit{vs.} Machine-learning Classification''.
In this experiment, VGG16 input patches were defined as superpixels, and \system six last layers were added at the end of the network topology.
Results indicate \system with InceptionV3 and ResNet outperformed VGG16 for every mean value by a small margin.
Therefore, we set ResNet as \system underlying CNN in the area quantification experiment.\\

\begin{figure}[!htb]
\centering
\includegraphics[scale=1.07]{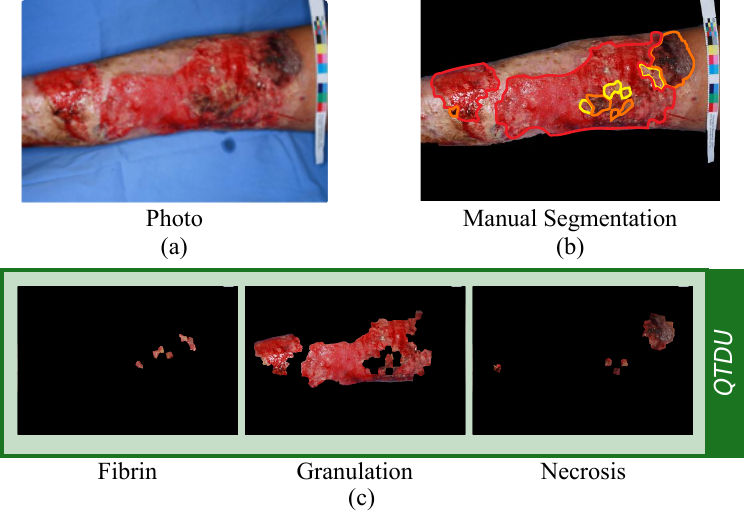}
\caption{Examples of 
(a)~wound photograph,
(b)~manual segmentation, and 
(c)~\system pixelwise area quantification.}
\label{fig:quantification}
\end{figure}

\noindent
\textbf{Pixelwise area quantification.}
Five images were sep\-a\-rat\-ed from the original \dataset dataset of 217 (minus 40) images, and two experts manually segmented them according to the four-color class model.
Figures~\ref{fig:quantification}(a--b) show a selected image and its segmentation in the ImageJ\footnote{\url{https://imagej.nih.gov/ij/}} tool.
Accordingly, we performed a holdout evaluation where the new photos were evaluated as the testing fold.
Next, we matched the pixels of \system with ResNet segmentation (Figure~\ref{fig:quantification}(c)) to the manually annotated pixels.
The \system average accuracy was $93.56\%$, which is $4.2\%$ higher than the best result reported by previous machine-learning studies, \textit{i.e.}, $89.87\%$~(See Section~\ref{sec:introduction}).
The Mean Absolute Error (MAE) was calculated for the segmented images counting the unmatched pixels and dividing the result by the number of wounded pixels.
The average MAE ratio was $0.089$ with a $0.015$ variance.

\noindent
\textbf{Discussion.}
\system outperformed existing three-stage approaches by significant margins in the segmentation of ulcers in lower limbs.
In particular, \system was up to $7.3\%$ (Sensitivity) and $8.7\%$ (Specificity) better than the best carefully constructed machine-learning combo: SLIC superpixels combined with MPEG-7 Scalable Color extractor, PCA reducer with Kaiser-Guttman criterion, and a tuned RandomForest classifier.
Such gains were corroborated by F1-Score confusion matrices, which indicate \system achieved top performances for individual classes of wounded tissues.
Although no significant differences were observed between InceptionV3 and ResNet as \system underlying CNNs, ResNet outperformed machine-learning approaches within stronger significance levels.
On the other hand, InceptionV3 was up $41\%$ faster to train than ResNet.
Results also indicate \system is flexible enough to use any patch-based CNN, \textit{e.g.}, VGG16, InceptionV3, or ResNet, whereas overall accuracy may be affected by the choice.
Therefore, the underlying CNN can be set towards optimizing either performance or resources.
Last, but not least, experiments showed \system with ResNet reached an average error of only $0.089$ in comparison to human-conducted segmentation.

\section{Conclusions}\label{sec:conclusion}

This manuscript presented a superpixel-driven deep learning approach for the segmentation of wounded tissues within dermatological ulcers.
The method, called \system, is a divide-and-conquer approach designed as a modular architecture.
\system takes advantage of superpixels for raw wound segmentation and uses coupled CNNs for performing feature extraction and tissue classification.
As a side-effect, \system disregards superpixels' spatial location in the original image.
We validated our proposal using a set of ulcer photos, whose pixels were labeled according to four tissue types.
\system advances in comparison to existing machine-learning-based approaches include
\textit{(i)}~bypassing the feature extraction step, and
\textit{(ii)}~providing a segmentation ``mask'' by joining labeled superpixels of the same class.

Empirical evaluations over 179,572 superpixels divided into four classes indicated \system efficiently segmented wounded tissues and outperformed fine-tuned machine-learning strategies in up to $7.3\%$, $8.7\%$, and $3.6\%$ regarding Sensitivity, Specificity, and AUC, respectively.
Additionally, unlike existing machine-learning approaches, \system was able to enhance the hit-ratio of every type of wounded tissues, simultaneously.
A ranking test also indicated \system outperformed every one of the eight machine-learning competitors for the classification of wound images within strong significant levels, whereas no significant differences were observed for \system with underlying CNNs InceptionV3 and ResNet.

Experiments indicated other underlying CNNs, such as VGG16, can be seamlessly used as \system parameterization, whereas the accuracy and running time of the proposal are influenced by that decision.
For instance, results showed ResNet average predictions were slightly higher than VGG16 and InceptionV3, but ResNet required more training and labeling time in comparison to competing CNNs.
Accordingly, \system can be parameterized towards performance or available resources, depending on the expert requirements.
The final experiment showed \system reached a $93.56\%$ accuracy with a merely $0.089$ Mean Absolute Error ratio when compared to a manual human segmentation.
Such findings reinforce \system can segment ulcers in lower limbs, and delimit the pixelwise area within wounded tissues.

As future work, we are designing a protocol to increase \dataset dataset so that it can become a benchmark for comparing wound analysis methods.
Additionally, we intend to extend \system by exploring the impact of several data augmentation strategies over superpixels to design a feature extractor module to be coupled into a content-based image retrieval tool for dermatological wounds.

\section{References}
\bibliographystyle{plain}
\bibliography{main}

\begin{thebibliography}{10}
\expandafter\ifx\csname url\endcsname\relax
  \def\url#1{\texttt{#1}}\fi
\expandafter\ifx\csname urlprefix\endcsname\relax\def\urlprefix{URL }\fi
\expandafter\ifx\csname href\endcsname\relax
  \def\href#1#2{#2} \def\path#1{#1}\fi

\bibitem{Gillies2015}
R.~J. Gillies, P.~E. Kinahan, H.~Hricak, {Radiomics: images are more than
  pictures, they are data}, Radiology 278~(2) (2015) 563--577.

\bibitem{Yu2017}
L.~Yu, H.~Chen, Q.~Dou, J.~Qin, P.~Heng, Automated melanoma recognition in
  dermoscopy images via very deep residual networks, IEEE Transactions on
  Medical Imaging 36~(4) (2017) 994--1004.
\newblock \href {http://dx.doi.org/https://doi.org/10.1109/TMI.2016.2642839}
  {\path{doi:https://doi.org/10.1109/TMI.2016.2642839}}.

\bibitem{Deserno2018}
S.~Kamath, E.~Sirazitdinova, T.~M. Deserno, Machine learning for mobile wound
  assessment, in: Imaging Informatics for Healthcare, Research, and
  Applications, Vol.~1, International Society for Optics and Photonics, 2018,
  pp. 1--8.
\newblock \href {http://dx.doi.org/https://doi.org/10.1117/12.2293704}
  {\path{doi:https://doi.org/10.1117/12.2293704}}.

\bibitem{Blanco2016}
G.~Blanco, M.~V.~N. Bedo, M.~T. Cazzolato, L.~F.~D. Santos, A.~E.~S. Jorge,
  C.~Traina{ }Jr., P.~M. Azevedo-Marques, A.~J.~M. Traina, {A Label-Scaled
  Similarity Measure for Content-Based Image Retrieval}, in: IEEE International
  Symposium on Multimedia, IEEE, 2016, pp. 20--25.
\newblock \href {http://dx.doi.org/http://dx.doi.org/10.1109/ISM.2016.0014}
  {\path{doi:http://dx.doi.org/10.1109/ISM.2016.0014}}.

\bibitem{Zahia2018}
S.~Zahia, D.~Sierra-Sosa, B.~Garcia-Zapirain, A.~Elmaghraby, Tissue
  classification and segmentation of pressure injuries using convolutional
  neural networks, Computer Methods and Programs in Biomedicine 159 (2018)
  51--58.
\newblock \href {http://dx.doi.org/https://doi.org/10.1016/j.cmpb.2018.02.018}
  {\path{doi:https://doi.org/10.1016/j.cmpb.2018.02.018}}.

\bibitem{Litjens2017}
G.~Litjens, T.~Kooi, B.~E. Bejnordi, A.~A.~A. Setio, F.~Ciompi, M.~Ghafoorian,
  J.~A. Van Der~Laak, B.~Van~Ginneken, C.~I. S{\'a}nchez, A survey on deep
  learning in medical image analysis, Medical Image Analysis 42 (2017) 60--88.
\newblock \href {http://dx.doi.org/https://doi.org/10.1016/j.media.2017.07.005}
  {\path{doi:https://doi.org/10.1016/j.media.2017.07.005}}.

\bibitem{Seixas2015}
J.~L. Seixas, S.~Barbon, R.~G. Mantovani, Pattern recognition of lower member
  skin ulcers in medical images with machine learning algorithms, in: IEEE
  International Symposium on Computer-Based Medical Systems, IEEE, 2015, pp.
  50--53.

\bibitem{Mukherjee2014}
R.~Mukherjee, D.~Manohar, D.~K. Das, A.~Achar, A.~Mitra, C.~Chakraborty,
  Automated tissue classification framework for reproducible chronic wound
  assessment, BioMed Research International 2014.
\newblock \href {http://dx.doi.org/http://dx.doi.org/10.1155/2014/851582}
  {\path{doi:http://dx.doi.org/10.1155/2014/851582}}.

\bibitem{Kavitha2017}
I.~Kavitha, S.~Suganthi, S.~Ramakrishnan, {Analysis of Chronic Wound Images
  Using Factorization Based Segmentation and Machine Learning Methods}, in:
  Proceedings of International Conference on Computational Biology and
  Bioinformatics, ACM, 2017, pp. 74--78.
\newblock \href {http://dx.doi.org/https://doi.org/10.1145/3155077.3155092}
  {\path{doi:https://doi.org/10.1145/3155077.3155092}}.

\bibitem{Pereyra2014}
S.~M. Pereira, M.~A.~C. Frade, R.~M. Rangayyan, P.~M. Azevedo-Marques,
  {Classification of color images of dermatological ulcers}, IEEE Journal of
  Biomedical and Health Informatics 17~(1) (2013) 136--142.
\newblock \href {http://dx.doi.org/10.1109/TITB.2012.2227493}
  {\path{doi:10.1109/TITB.2012.2227493}}.

\bibitem{Veredas2010}
F.~Veredas, H.~Mesa, L.~Morente, Binary tissue classification on wound images
  with neural networks and bayesian classifiers, IEEE Transactions on Medical
  Imaging 29~(2) (2010) 410--427.
\newblock \href {http://dx.doi.org/https://doi.org/10.1109/TMI.2009.2033595}
  {\path{doi:https://doi.org/10.1109/TMI.2009.2033595}}.

\bibitem{Chino2018}
D.~Y.~T. Chino, L.~C. Scabora, M.~T. Cazzolato, A.~E.~S. Jorge, C.~Traina{
  }Jr., A.~J.~M. Traina, {ICARUS: Retrieving Skin Ulcer Images through
  Bag-of-Signatures}, in: IEEE International Symposium on Computer-Based
  Medical Systems, IEEE, 2018, pp. 82--87.
\newblock \href {http://dx.doi.org/https://doi.org/10.1109/CBMS.2018.00022}
  {\path{doi:https://doi.org/10.1109/CBMS.2018.00022}}.

\bibitem{Achanta2012}
R.~Achanta, A.~Shaji, K.~Smith, A.~Lucchi, P.~Fua, S.~S{\"u}sstrunk, {SLIC
  superpixels compared to state-of-the-art superpixel methods}, IEEE
  Transactions on Pattern Analysis and Machine Intelligence 34~(11) (2012)
  2274--2282.
\newblock \href {http://dx.doi.org/https://doi.org/10.1109/TPAMI.2012.120}
  {\path{doi:https://doi.org/10.1109/TPAMI.2012.120}}.

\bibitem{Dorileo2010}
E.~A.~G. Dorileo, M.~A.~C. Frade, R.~M. Rangayyan, P.~M. Azevedo-Marques,
  Segmentation and analysis of the tissue composition of dermatological ulcers,
  in: Canadian Conference of Electrical and Computer Engineering, IEEE, 2010,
  pp. 1--4.
\newblock \href
  {http://dx.doi.org/http://dx.doi.org/10.1109/CCECE.2010.5575143}
  {\path{doi:http://dx.doi.org/10.1109/CCECE.2010.5575143}}.

\bibitem{Khalil2019}
A.~Khalil, M.~Elmogy, M.~Ghazal, C.~Burns, A.~El-Baz, {Chronic Wound Healing
  Assessment System Based on Different Features Modalities and Non-Negative
  Matrix Factorization (NMF) Feature Reduction}, IEEE Access 7 (2019)
  80110--80121.
\newblock \href {http://dx.doi.org/10.1109/ACCESS.2019.2923962}
  {\path{doi:10.1109/ACCESS.2019.2923962}}.

\bibitem{Esteva2017}
A.~Esteva, B.~Kuprel, R.~A. Novoa, J.~Ko, S.~M. Swetter, H.~M. Blau, S.~Thrun,
  Dermatologist-level classification of skin cancer with deep neural networks,
  Nature 542~(7639) (2017) 115.
\newblock \href {http://dx.doi.org/https://doi.org/10.1038/nature21056}
  {\path{doi:https://doi.org/10.1038/nature21056}}.

\bibitem{Yuexiang2018}
Y.~Li, L.~Shen, Skin lesion analysis towards melanoma detection using deep
  learning network, Sensors (Basel) 18~(2).
\newblock \href {http://dx.doi.org/https://doi.org/10.3390/s18020556}
  {\path{doi:https://doi.org/10.3390/s18020556}}.

\bibitem{Wang2015}
C.~Wang, X.~Yan, M.~Smith, K.~Kochhar, M.~Rubin, S.~M. Warren, J.~Wrobel,
  H.~Lee, A unified framework for automatic wound segmentation and analysis
  with deep convolutional neural networks, in: International Conference of the
  IEEE Engineering in Medicine and Biology Society, IEEE, 2015, pp. 2415--2418.
\newblock \href {http://dx.doi.org/10.1109/EMBC.2015.7318881}
  {\path{doi:10.1109/EMBC.2015.7318881}}.

\bibitem{Simonyan2014}
K.~Simonyan, A.~Zisserman, Very deep convolutional networks for large-scale
  image recognition, arXiv preprint arXiv:1409.1556.

\bibitem{Russakovsky2015}
O.~Russakovsky, J.~Deng, H.~Su, J.~Krause, S.~Satheesh, S.~Ma, Z.~Huang,
  A.~Karpathy, A.~Khosla, M.~Bernstein, A.~C. Berg, L.~Fei-Fei, Imagenet large
  scale visual recognition challenge, International Journal of Computer Vision
  115~(3) (2015) 211--252.
\newblock \href {http://dx.doi.org/https://doi.org/10.1007/s11263-015-0816-y}
  {\path{doi:https://doi.org/10.1007/s11263-015-0816-y}}.

\bibitem{He2016}
K.~He, X.~Zhang, S.~Ren, J.~Sun, Deep residual learning for image recognition,
  in: Proceedings of IEEE Conference on Computer Vision and Pattern
  Recognition, 2016, pp. 770--778.
\newblock \href {http://dx.doi.org/https://doi.org/10.1109/CVPR.2016.90}
  {\path{doi:https://doi.org/10.1109/CVPR.2016.90}}.

\bibitem{Szegedy2016}
C.~Szegedy, V.~Vanhoucke, S.~Ioffe, J.~Shlens, Z.~Wojna, Rethinking the
  inception architecture for computer vision, in: IEEE Conference on Computer
  Vision and Pattern Recognition, 2016, pp. 2818--2826.
\newblock \href {http://dx.doi.org/https://doi.org/10.1109/CVPR.2016.308}
  {\path{doi:https://doi.org/10.1109/CVPR.2016.308}}.

\bibitem{Goyal2018}
M.~Goyal, N.~D. Reeves, A.~K. Davison, S.~Rajbhandari, J.~Spragg, M.~H. Yap,
  {Dfunet: Convolutional neural networks for diabetic foot ulcer
  classification}, IEEE Transactions on Emerging Topics in Computational
  Intelligence (2018) 1--12\href
  {http://dx.doi.org/https://doi.org/10.1109/TETCI.2018.2866254}
  {\path{doi:https://doi.org/10.1109/TETCI.2018.2866254}}.

\bibitem{Nejati2018}
H.~Nejati, H.~A. Ghazijahani, M.~Abdollahzadeh, T.~Malekzadeh, N.~M. Cheung,
  K.~H. Lee, L.~L. Low, Fine-grained wound tissue analysis using deep neural
  network, in: IEEE International Conference on Acoustics, Speech and Signal
  Processing, IEEE, 2018, pp. 1010--1014.
\newblock \href {http://dx.doi.org/10.1109/ICASSP.2018.8461927}
  {\path{doi:10.1109/ICASSP.2018.8461927}}.

\bibitem{Shin2016}
S.~Hoo-Chang, H.~R. Roth, M.~Gao, L.~Lu, Z.~Xu, I.~Nogues, J.~Yao, D.~Mollura,
  R.~M. Summers, {Deep convolutional neural networks for computer-aided
  detection: CNN architectures, dataset characteristics and transfer learning},
  IEEE Transactions on Medical Imaging 35~(5) (2016) 1285--1298.
\newblock \href {http://dx.doi.org/https://doi.org/10.1109/TMI.2016.2528162}
  {\path{doi:https://doi.org/10.1109/TMI.2016.2528162}}.

\bibitem{Han2018}
S.~Han, M.~Kim, W.~Lim, G.~Park, I.~Park, S.~Chang, Classification of the
  clinical images for benign and malignant cutaneous tumors using a deep
  learning algorithm, Journal of Investigative Dermatology 138~(7) (2018)
  1529--1538.
\newblock \href {http://dx.doi.org/https://doi.org/10.1016/j.jid.2018.01.028}
  {\path{doi:https://doi.org/10.1016/j.jid.2018.01.028}}.

\bibitem{Youssef2018}
A.~Youssef, D.~Bloisi, M.~Muscio, A.~Pennisi, D.~Nardi, A.~Facchiano, Deep
  convolutional pixel-wise labeling for skin lesion image segmentation, in:
  IEEE International Symposium on Medical Measurements and Applications, 2018,
  pp. 1--6.
\newblock \href
  {http://dx.doi.org/https://doi.org/10.1109/10.1109/MeMeA.2018.8438669}
  {\path{doi:https://doi.org/10.1109/10.1109/MeMeA.2018.8438669}}.

\bibitem{Sonka2014}
M.~Sonka, V.~Hlavac, R.~Boyle, Image processing, analysis, and machine vision,
  Cengage Learning, 2014.

\bibitem{Luo2018}
W.~Luo, M.~Yang, Fast skin lesion segmentation via fully convolutional network
  with residual architecture and {CRF}, in: International Conference on Pattern
  Recognition, 2018, pp. 1438--1443.
\newblock \href {http://dx.doi.org/10.1109/ICPR.2018.8545571}
  {\path{doi:10.1109/ICPR.2018.8545571}}.

\bibitem{Krizhevsky2012}
A.~Krizhevsky, I.~Sutskever, G.~E. Hinton, Imagenet classification with deep
  convolutional neural networks, in: Advances in Neural Information Processing
  Systems, 2012, pp. 1097--1105.

\bibitem{Szegedy2015}
C.~Szegedy, W.~Liu, Y.~Jia, P.~Sermanet, S.~Reed, D.~Anguelov, V.~Vanhoucke,
  A.~Rabinovich, Going deeper with convolutions, in: IEEE Conference on
  Computer Vision and Pattern Recognition, 2015, pp. 1--9.
\newblock \href {http://dx.doi.org/https://doi.org/10.1109/CVPR.2015.7298594}
  {\path{doi:https://doi.org/10.1109/CVPR.2015.7298594}}.

\bibitem{Yap2018}
J.~Yap, W.~Yolland, P.~Tschandl, Multimodal skin lesion classification using
  deep learning, Experimental Dermatology 27~(11) (2018) 1261--1267.
\newblock \href {http://dx.doi.org/https://doi.org/10.1111/exd.13777}
  {\path{doi:https://doi.org/10.1111/exd.13777}}.

\bibitem{Kawahara2016}
J.~Kawahara, A.~BenTaieb, G.~Hamarneh, Deep features to classify skin lesions,
  in: International Symposium on Biomedical Imaging, IEEE, 2016, pp.
  1397--1400.
\newblock \href {http://dx.doi.org/https://doi.org/10.1109/ISBI.2016.7493528}
  {\path{doi:https://doi.org/10.1109/ISBI.2016.7493528}}.

\bibitem{Van2014}
L.~V.~D. Maaten, {Accelerating t-SNE using tree-based algorithms}, Journal of
  Machine Learning Research 15~(1) (2014) 3221--3245.

\bibitem{Mello2018}
R.~F. Mello, M.~A. Ponti, {Machine Learning: A Practical Approach on the
  Statistical Learning Theory}, Springer, 2018.

\bibitem{Peres2005}
P.~R. Peres-Neto, D.~A. Jackson, K.~M. Somers, {How many principal components?
  Stopping rules for determining the number of non-trivial axes revisited},
  Computational Statistics \& Data Analysis 49~(4) (2005) 974--997.
\newblock \href {http://dx.doi.org/https://doi.org/10.1016/j.csda.2004.06.015}
  {\path{doi:https://doi.org/10.1016/j.csda.2004.06.015}}.

\bibitem{Hinton2012}
G.~E. Hinton, N.~Srivastava, A.~Krizhevsky, I.~Sutskever, R.~R. Salakhutdinov,
  \href{https://arxiv.org/pdf/1207.0580}{Improving neural networks by
  preventing co-adaptation of feature detectors}, arXiv preprint:1207.0580.
\newline\urlprefix\url{https://arxiv.org/pdf/1207.0580}

\bibitem{Demsar2006}
J.~Demsar, Statistical comparisons of classifiers over multiple data sets,
  Journal of Machine Learning Research 7 (2006) 1--30.

\bibitem{Garcia2008}
S.~Garcia, F.~Herrera, {An extension on ``Statistical comparisons of
  classifiers over multiple data sets'' for all pairwise comparisons}, Journal
  of Machine Learning Research 9 (2008) 2677--2694.

\end{thebibliography}

\end{document}